# CORRELAÇÃO DE LONGA-DURAÇÃO E ANALÍSE BAYESIANA DA EVOLUÇÃO TEMPORAL DOS TERREMOTOS AO LONGO DA FALHA DE SAMAMBAIA, NORDESTE DO BRASIL

**Daniel Brito de Freitas**
NECMAP, Instituto de Educação, Ciência e Tecnologia do Rio Grande do Norte, Campus João Câmara, Bolsista Programa de Apoio Institucional à Pesquisa.
daniel.brito@ifrn.edu.br

**George Sand França**
Observatório Sismológico da Universidade de Brasília. georgesand@unb.br

**Leandro Luiz da Silva Perreira**
NECMAP, Instituto de Educação, Ciência e Tecnologia do Rio Grande do Norte, Campus João Câmara, Bolsista Programa de Apoio Institucional à Pesquisa.

**Leandro da Silva Pedro**
NECMAP, Instituto de Educação, Ciência e Tecnologia do Rio Grande do Norte, Campus João Câmara, Bolsista Programa de Apoio Institucional à Pesquisa.

**Maria da Glória Nascimento Atanazio**
NECMAP, Instituto de Educação, Ciência e Tecnologia do Rio Grande do Norte, Campus João Câmara, Bolsista Programa de Apoio Institucional à Pesquisa.
ggloriaglorinha@hotmail.com

**Carlos da SilvaVilar**
Instituto de Física da Universidade Federal da Bahia. vilar@ufba.br

**RESUMO**
Uma aproximação Bayesiana é adotada para analisar a seqüência de eventos sísmicos e suas magnitudes próximo a João Câmara que ocorreram principalmente de 1983 a 1998 ao longo da Falha de Samambaia. Neste trabalho, nós escolhemos um modelo Bayesiano para o processo de tempo de ocorrência condicional sobre os valores de magnitudes observadas seguindo o mesmo procedimento proposto por Stavrakakis e Tselentis (1987). Os parâmetros do modelo são determinados sobre uma base de informações físicas e históricas. Nós geramos uma amostra *a posteriori* a partir de sua distribuição através de uma variante do algoritmo Metropolis-Hastings. Nós usamos os resultados em uma variedade de aplicações, incluindo a construção de um intervalo de confiança para a intensidade condicional do processo como uma função de tempo, como também, uma distribuição *a posteriori* como uma função da má ocorrência por unidade de tempo.

**PALAVRAS-CHAVE:** terremotos, estatística Bayesiana, eventos sísmicos, falha de Samambaia.

# LONG-RANGE CORRELATION AND BAYESIAN ANALYSIS OF TIME EVOLUTION OF EARTHQUAKES ALONG THE SAMAMBAIA FAULT, NORTHEAST BRAZIL

**RESUMO (segunda língua)**
A Bayesian approach is adopted to analyze the sequence of seismic events and their magnitudes near João Câmara which occurred mainly from 1983 to 1998 along the Samambaia fault. In this work, we choose a Bayesian model for the process of occurrence times conditional on the observed magnitude values following the same procedure suggested by Stavrakakis and Tselentis (1987). The model parameters are determined on the basis of historical and physical information. We generate posterior samples from the joint posterior distribution of the model parameters by using a variant of the Metropolis-Hastings algorithm. We use the results in a variety of ways, including the construction of pointwise posterior confidence bands for the conditional intensity of the point process as a function of time, as well as, a posterior distribuition as a function of the mean of occurrence per unit time.

**PALAVRAS-CHAVE (segunda língua):** earthquakes, Bayesian statistics, seismic events, Samambaia fault.

# CORRELAÇÃO DE LONGA-DURAÇÃO E ANALÍSE BAYESIANA DA EVOLUÇÃO TEMPORAL DOS TERREMOTOS AO LONGO DA FALHA DE SAMAMBAIA, NORDESTE DO BRASIL

## INTRODUÇÃO

O nordeste brasileiro é, segundo recentes estudos sobre a sismicidade histórica e instrumental, uma das áreas de maior atividade sísmica da América do Sul. Estas atividades tem ocorrido frequentemente ao redor da bacia Potiguar. A sismicidade é classificada por eventos de magnitude moderada < 5.2 na escala Richter e eventos de alta magnitude > 7.0, também conhecidos como eventos paleosísmicos. Nosso estudo se concentra na área epicentral de João Câmara, onde o conjunto de terremotos compõe uma amostra de mais de 1000 eventos ocorridos principalmente de 1983 a 1998 ao longo da falha de Samambaia. Segundo Bezerra et al. (2007), apenas 14 destes eventos possuem magnitudes maiores que 4.0 e apenas dois com magnitudes maiores que 5.0. Para termos uma idéia destas magnitudes e como elas são raras, basta compararmo-nas com a intensidade dos recentes terremotos que ocorreram na região de João Câmara nos dias 9 e 11 de janeiro deste ano que atingiram a marca de 3.8, segundo dados do Observatório Sismológico da UnB. Nós também agregamos a esta sequência de eventos os dados mais recentes dos últimos terremotos.

Os terremotos são alguns dos mais relevantes paradigmas da conhecida criticalidade auto-organizada, introduzida por Bak, Tang e Wiesenfeld (1987). No contexto de sistemas com falha, como é o caso da Falha de Samambaia, isto representa um fenômeno com complexidade espaço-temporal, investigado através da deformação e ruptura repentina de algumas partes da crosta terrestre controlada pelo movimento convectivo no manto, isto é, a radiação de energia na forma de ondas sísmicas. Em particular, a Falha de Samambaia é, segundo Silva et al. (2006), um exemplo característico de sistema complexo auto-organizado na natureza. Assim como inúmeros estudos a respeito dos sistemas complexos na natureza, a investigação sobre o comportamento dos terremotos passa pela questão da previsibilidade, ou seja, que parâmetros e variáveis extraídas dos dados sismológicos, ou outra espécie de registro, são indicadores que possam prever determinado evento? Vale ressaltar que o termo evento não se refere a uma data precisa, mas se no futuro este evento poderá ocorrer e qual o grau de confiabilidade estatística desta previsão.

Este conjunto de evidências foi observado no dia 30 de novembro de 1986 na região de João Câmara no Estado do Rio Grande do Norte. Para a comunidade cientifica este evento pode ser considerado um marco científico, social e histórico para o Brasil. Na madrugada deste dia, um terremoto atingiu a magnitude 5.1 na escala Richter, colocando a população local em pânico e as autoridades estaduais e federais em alerta. Como relatou Amaral (2000), ele constitui um marco científico porque despertou o interesse da comunidade geológica para o estudo da atividade neotectônica no Nordeste, até então relegada a segundo plano; do ponto de vista social, porque tiveram que ser repensadas e modificadas as maneiras de construções das edificações na região afetada; e relativo a história pois, pela primeira vez, a comunidade científica brasileira presenciou e estudou os efeitos físicos de uma sequência de tremores, até então só conhecidos através da literatura ou relatada em outras partes do mundo como o Chile e os Estados Unidos. Além dos impactos científico, social e histórico relatados por Amaral (2000), podemos citar uma outra vertente, o *impacto psicológico*. Logo após os terremotos de 86, várias famílias João Camarense, abaladas psicologicamente com a fragilidade de suas casas e incertas perante o futuro de João Câmara, abandonaram ou venderam a baixíssimo preços suas casas e terrenos, afetando drasticamente a economia e a ordem social desta cidade. Talvez esta tenha sido a pior conseqüência após os terremotos. Ainda hoje, os moradores relatam que o progresso e desenvolvimento foi dramaticamente afetado pelos abalos de 86, fato que, segundo eles, tornou uma promissora cidade ao status de uma cidade que a qualquer momento pode voltar a estaca zero como em 86. Relatos da população de João Câmara revelam que, mesmo depois de mais de 20 anos, os terremotos ainda são vistos como um inimigo oculto.

Neste sentido, através de um procedimento inédito e robusto que une a informação amostral (dados sismológicos) e a informação subjetiva (opiniões, experiências, etc.), podemos extrair informações de onde a maioria dos cientistas colocaria em segundo plano ou literalmente ignoraria em sua pesquisa, a informação cultural. Isto, em síntese, abrirá um canal entre o conhecimento cientifico e as práticas sociais que refletem diretamente no cenário educacional. Recentemente, pesquisas desta natureza vem sendo realizadas no Ceará com os *Profetas da Seca* (Andrade 2007).

Finalmente, vale ressaltar que o trabalho questões de caráter físico, ou seja, parâmetros e variáveis que controlam a dinâmica espaço-temporal dos abalos sísmicos desta região.

**ESTADO DA ARTE**

Segundo Wilson (1966), de acordo com a teoria clássica das placas, os continentes são internamente estáveis e os movimentos tectônicos estão concentrados principalmente na linha que divide as placas. Contudo, como citou Amaral (2000), a freqüência e distribuição de terremotos em seus interiores, assim como as medidas geodésicas, evidenciando deslocamentos intraplacas, tanto verticais como horizontais em várias partes do mundo, levaram a reconhecer que muitas regiões da superfície terrestre, consideradas como estáveis, estão sujeitas em maior ou menor grau a essas vibrações. Ao longo de sua história geológica, a Terra vem experimentando as mais diversas transformações que repercutem na sua superfície em forma de movimentos epirogenéticos, orogenéticos, deslocamentos de placas, os quais, associados ao equilíbrio isostático, produzem as mais diversas formas estruturais.

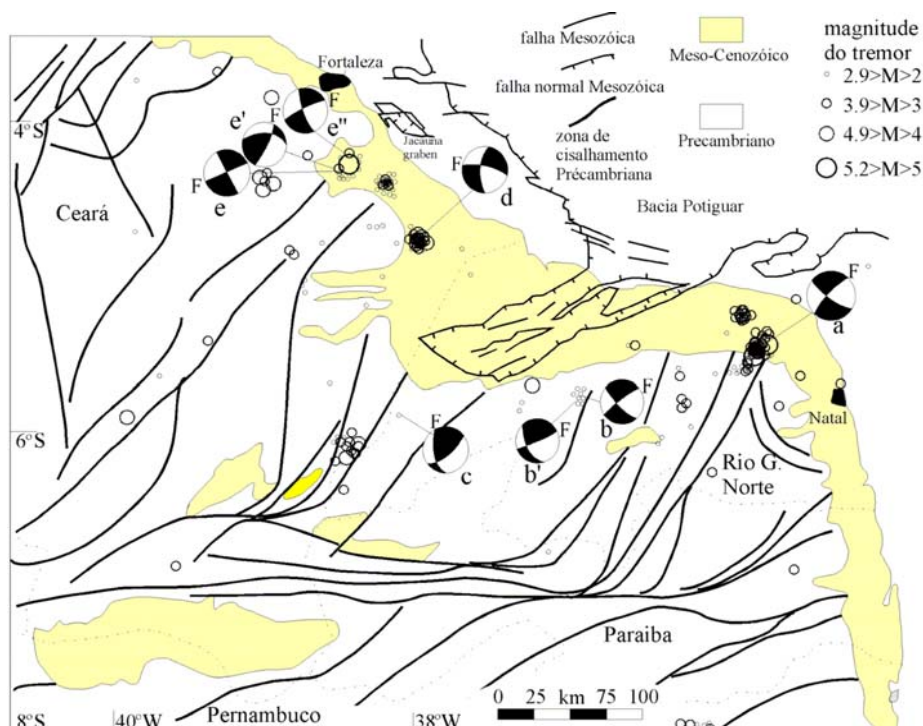

**Figura. 1: Sismicidade e mecanismo focais no Nordeste Oriental. A letra (a) representa a região de João Câmara. Figura modificada de Berrocal *et al.* (1984) e Ferreira *et al.* (1987)**

Como mencionou Silva et al. (2006), existem alguns ingredientes físicos que controlam a dinâmica dos terremotos tornando-os fenômenos fora do equilíbrio termodinâmico e, por sua vez, dirigidos por interações de longo alcance. São eles:

1) O mecanismo de deslocamento relativo das falhas;
2) A irregularidade na superfície das placas tectônicas e o espaço fragmentado entre as estruturas internas;
3) O mecanismo de que aciona os terremotos dado pela combinação de irregularidades das falhas e a distribuição de fragmentos entre elas;
4) E por se tratar de um sistema fora do equilíbrio revelada pelo surgimento de forte caudas na função de distribuição das magnitudes.

Segundo de Freitas (2009), forças de interação de longo alcance estão relacionadas com efeitos de *memória de longa duração* apresentando uma estrutura fractal que impede que modelos clássicos, válidos apenas no regime de memória de curta duração, sejam adequados para explicar a dinâmica dos terremotos. A memória de longa duração está presente em um dado fenômeno se uma determinada escala de tempo se repete ao longo do tempo (Souza et al. 2006). Um exemplo prático é o ciclo de atividade solar que se repete a cada 11 anos (de Freitas & De Medeiros 2009). Neste sentido, se encontramos indícios deste tipo de memória podemos inferir sobre possíveis periodicidades, ou seja, se determinado evento vai se repetir no futuro.

Para reforçar este ponto, Silva et al. (2006; 2007) mostraram que a função de distribuição do número de terremotos como uma função da magnitude presente na Falha de Samambaia representa um sistema fora do equilíbrio e apresenta memória interna de longa duração. Neste sentido, modelos clássicos baseados na Mecânica Estatística de Boltzmamm-Gibbs, que consideram, a priori, que os terremotos não apresentam memória ou apenas memória de curta duração, falham no diagnóstico da previsão destes fenômenos.

Estudos recentes sobre a sismologia do globo terrestre através do Global Seismic Hazard Assessment Program (GSHAP) apontam que João Câmara pode ser considerada uma área de atividade sísmica moderada (Takeya 1992, Shedlock et al. 2000). A região de João Câmara é a principal área sísmica da região e do país, com o maior acervo de dados instrumentais. A evolução da sismicidade, também denominada de série temporal dos abalos sísmicos, permitiu o registro da nucleação e desenvolvimento de uma falha, denominada por Takeya et al. (1989) de Falha Sísmica de Samambaia. Uma importante propriedade geológica desta falha, apontado por Ferreira et al. (1987), é que a mesma não coincide, inclusive na sua direção, com as estruturas pré-cambrianas até então mapeadas. Isto torna esta região particular em relação a outras falhas que se encontram no Brasil. A Figura 1 nos mostra em detalhes o quadro sismológico de uma parte do Nordeste brasileiro enfatizando a origem geológica das falhas, assim como a escala de magnitude dos terremotos observados nesta área.

A grande maioria dos artigos da literatura tratam o comportamento espaço-temporal de fenômenos segundo a ótica da informação amostral (dados científicos). Existe uma corrente de físicos que utilizam modelos estatísticos baseados no Teorema de Bayes (Andrade & Ohagan 2006) que se opõe à visão dos frequentistas. Segundo o Teorema de Bayes (Bayes 1763), a probabilidade de 50% que a moeda caia com a face cara ou coroa é apenas um estado de conhecimento atual, ou seja, uma situação onde não podemos privilegiar a cara em detrimento da coroa. Esta informação é resultado da simetria do estado, por isso admitimos a mesma probabilidade. Ainda segundo Bayes (1763), nós podemos condicionar ao sistema uma nova informação, por exemplo, a topologia da moeda que vai desequilibrar este estado equiprovável e gerar uma nova probabilidade para as faces da moeda.

Recentemente, Andrade (2007) empregou a Estatística Bayesiana para cruzar as informações dos dados de chuva da FUNCEME (Fundação Cearense de Meterologia) com as previsões dos Profetas da Seca. Segundo ele, na área de previsões climáticas, podem ser usada no tratamento tanto as informações científicas, obtidas por meio de estudos da atmosfera, imagens de satélites etc, quanto as informações subjetivas, obtidas pela observação da natureza no microssistema de interesse. Andrade (2007) obteve resultados que comprovaram que o ciclo de chuvas previstos pela FUNCEME, condicionada às informações dos profetas, mostraram informações mais precisas sobre as previsões das chuvas no interior do Estado do Ceará.

Existe pouca informação na literatura sobre o tema proposto pela nosso trabalho. Neste sentido, o procedimento que propomos para o tratamento de dados sismológicos através da Inferência Bayesiana representa um trabalho inédito e de profundo impacto no cenário da pesquisa científica sobre terremotos.

**METODOLOGIA**

Nossa pesquisa foi dividida em três passos. São eles:

1) Analisar estatisticamente os dados sismológicos referentes à Falha de Samambaia com janela temporal de 1983 a 1998, totalizando mais de 1000 eventos obtidos da estação brasileira WWSSN localizada em Brasília;

2) Catalogar, por meio de questionário, as informações da população local sobre os possíveis parâmetros naturais que eles associam aos terremotos;

3) Utilizar a Inferência Bayesiana para condicionar os dados sismológicos e as informações subjetivas e, em seguida, repetir o procedimento estatístico adotado no primeiro ponto e convoluir os resultados.

**Passo 1**
No tocante ao primeiro passo, procederemos da seguinte forma:

Uma série de observações apresenta memória de longa duração quando os valores observados em *lags* (defasagem temporal) distantes são correlacionados entre si, ou ainda, se o efeito de um evento em um dado instante pode ser detectado muitos *lags* depois. O objetivo deste passo é: verificar a existência de indicadores de memória de longa duração presentes nos dados sismológicos e calcular o coeficiente de autocorrelação dependente de escala, dado por

$$R(\theta, L) = \frac{\sum_{i=1}^{n-L-1}(\theta_{i+L} - \hat{\theta})(\theta_i - \hat{\theta})}{\sum_{i=1}^{n-1}(\theta_i - \hat{\theta})^2} \qquad \text{(equação 1)},$$

onde $\theta$ representa os dados sismológicos com uma função do tempo, $\hat{\theta}$ representa o valor médio desta variável e $L$ é o período que se repete ao longo da série temporal.

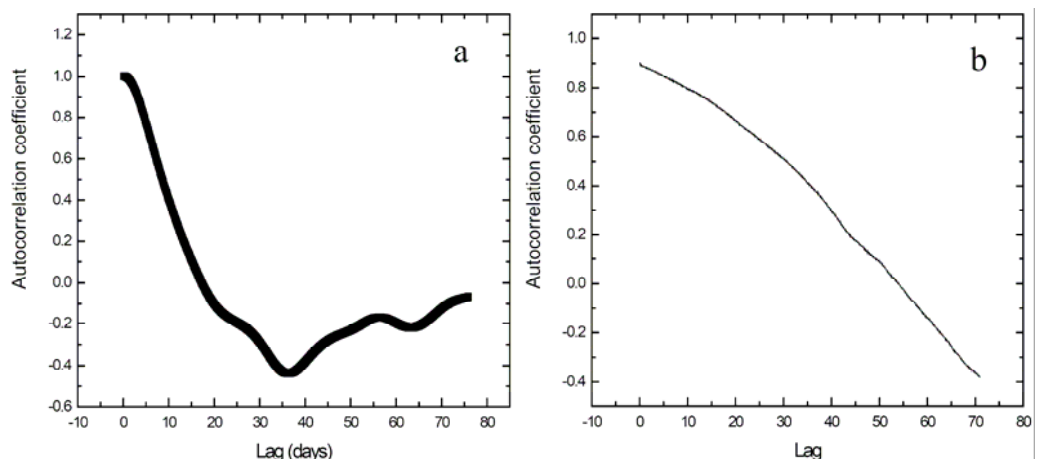

**Figura. 2: Exemplo do comportamento do coeficiente de autocorrelação em função do atraso (ou defasagem) am dias. Estas figuras foram extraídas de de Freitas et al. (2010).**

O comportamento deste coeficiente ao longo do tempo nos indicará a grandeza da memória do sistema. No caso da memória de curta duração, o coeficiente de autocorrelação decai de forma exponencial, tendendo a zero rapidamente como ilustrado na Figura 2(a). Entretanto, quando este coeficiente decai suavemente, a memória é de longa dura cão como ilustrado na Figura 2(b). Este coeficiente de autocorrelação esta diretamente ligado ao conhecido expoente de Hurst que mede o nível de memória do sistema.

No presente trabalho, utilizaremos diferentes estimadores deste expoente, dentre eles, a análise R/S clássica proposta por Hurst (1951). Para todos os estimadores utilizaremos dados embaralhados dentro de blocos contíguos ao longo de toda a série, por meio de uma janela móvel com o intento de identificar, em séries de retornos e de volatilidade de retornos, a ocorrência de previsibilidade devido à memória de longa dura cão e sua variabilidade ao longo do tempo (Souza *et al.* 2006), referentes à variabilidade da magnitude dos terremotos registrados na Falha de Samambaia denominada por $\theta$.

A metodologia utilizada nesta pesquisa foi inicialmente proposta por Hurst (1951) que formulou e derivou a relação empírica que fornece o que ficou conhecido como expoente de Hurst. Os valores deste expoente podem ser interpretados da seguinte forma: se estiver no intervalo de $0<H<0.5$ a série é antipersistente ou de memória de curta duração; se $H=0.5$ a série representa uma caminhada aleatória (random walk); por último, se $0.5<H<1.0$ a série é persistente ou de memória de longa duração. Para calcular o expoente de Hurst utilizaremos um procedimento semelhante ao proposto por Karagiannis (2002). Neste contexto, segue os diferentes métodos para estimar o expoente de Hurst empregados neste trabalho:

a) o método do valor absoluto;
b) o método da variância;
c) o método R/S;
d) o método do Periodograma;
e) o estimador Whittle;
f) o método da variância dos resíduos;
g) e o método de Abry-Veitch.

Neste contexto, o expoente de Hurst H será determinado com intervalo de confiança de 95%.

**Passo 2**

Neste passo analisamos criteriosamente as respostas de cada pessoa e estabeleceremos qual(is) parâmetro(s) foi(foram) dominante(s) nas respostas. A estes parâmetros atribuiremos a variável aleatória *X*. Esta variável pode ser um conjunto de parâmetros dados por $X = (x_1, x_2, ..., x_n)$.

**Passo 3**

Neste passo consideramos inicialmente que o verdadeiro valor do parâmetro $\theta$ é desconhecido, neste sentido, o objetivo é tentar reduzir este desconhecimento através da inclusão de outro parâmetro externo aos dados sismológicos obtidos no *Passo 2*. Além disso, a intensidade da incerteza a respeito de $\theta$ pode assumir diferentes graus. Do ponto de vista Bayesiano, estes diferentes graus de incerteza são representados através de modelos probabilísticos para $\theta$. Sendo assim, não existe nenhuma distinção entre quantidades observáveis e os parâmetros de um modelo estatístico, todos são considerados quantidades aleatórias. Deste modo, seguiremos com o procedimento:

a) neste primeiro ponto encontraremos a distribuição condicionada entre os dados sismológicos e as variáveis aleatórias determinadas no *Passo 2*. A informação que dispomos de *θ*, resumida através da probabilidade $p(\theta)$, pode ser ampliada pela inclusão da quantidade aleatória *X* relacionada com *θ*. Esta probilidade é extraída diretamente dos dados sismológicos. A distribuição amostral condicionada $p(\theta|x)$ define esta relação. A idéia de que após observar *X = x* a quantidade de informação sobre *θ* aumenta é bastante intuitiva e o teorema de Bayes é a regra de atualização utilizada para quantificar este aumento de informação,

$$p(\theta|x) = \frac{p(\theta,x)}{p(x)} = \frac{p(x|\theta)p(\theta)}{p(x)} \frac{p(x|\theta)p(\theta)}{\int p(\theta,x)d\theta} \qquad \text{(equação 2)},$$

em síntese *x* é o parâmetro dominante extraído do *Passo 2*.

Prosseguindo, utilizaremos a forma usual do Teorema de Bayes que determina a probabilidade de verossimilhança $l(\theta;x)$:

$$p(\theta|x) \propto l(\theta;x)p(\theta) \qquad \text{(equação 3)},$$

em outras palavras temos que

*distribuição a posteriori* $\propto$ *verossimilhança X distribuição a priori*

Caso mais de uma variável aleatória seja definida no anterior a equação 3 assumirá a seguinte forma:

$$p(\theta | x_1, x_2, ..., x_n) = \left[\prod_{i-1}^{n} l_i(\theta;x_i)\right] p(\theta) \qquad \text{(equação 4)};$$

b) calculada a distribuição a posteriori, prosseguimos com obtenção dos dados da magnitude dos terremotos condicionada às variáveis atribuídas no passo anterior. Ou seja, obteremos uma nova série temporal semelhante à série original de *θ*. Para isso, iremos recorrer ao Algoritmo de Metropolis-Hastings oriundo do Método de Monte-Carlo. Um método robusto amplamente utilizado pelos físicos para determinar valores esperados de propriedades de um sistema simulado. Com este método iremos inferir sobre a forma de $l(\theta;x)$;

c) neste último ponto, iremos comparar nosso modelo Bayesiano com o modelo de Poisson proposto por Stavrakakis e Tselentis (1987) e verificar como os mesmos se comportam para altas magnitudes.

## RESULTADOS E DISCUSSÕES

Nesta seção, é feita a estimação da memória de longo prazo variável no tempo, para a variabilidade dos dados sismológicos da falha de Samambaia, por meio de vários métodos, dentre eles, o coeficiente de autocorrelação e a análise R/S clássica, de Hurst (1951) e Mandelbrot (1972), calculada com janela móvel de aproximadamente 1000 observações e embaralhamento de dados dentro de blocos contíguos de dez observações, seguindo a abordagem empregada por Souza *et al.* (2006). Além deste tratamento, utilizaremos a Inferência Estatística para estudar o efeito das elevadas magnitudes sobre a distribuição cumulativa.

**Coeficiente de auto-correlação**

Em estatística, coeficiente de autocorrelação é uma medida que informa o quanto o valor de uma variável aleatória é capaz de influenciar seus vizinhos. Existem várias interpretações físicas e definições deste coeficiente mas, segundo a definição da estatística, o valor da autocorrelação está entre 1 (correlação perfeita) e -1, o que significa anti-correlação perfeita enquanto o valor "zero" significa total ausência de correlação.

A Figura 3 representa a série temporal dos terremotos ocorridos ao longo da Falha de Samambaia no período de 1983 a 1998. Este série revela as flutuações dos terremotos medidos pelos sismógrafos que foram catalogadas de forma descontínua, ou seja, a série não é igualmente espaçada. Isto pode provocar efeitos do tipo *gap* e, dependendo do tipo de método, pode ocasionar erros consideráveis.

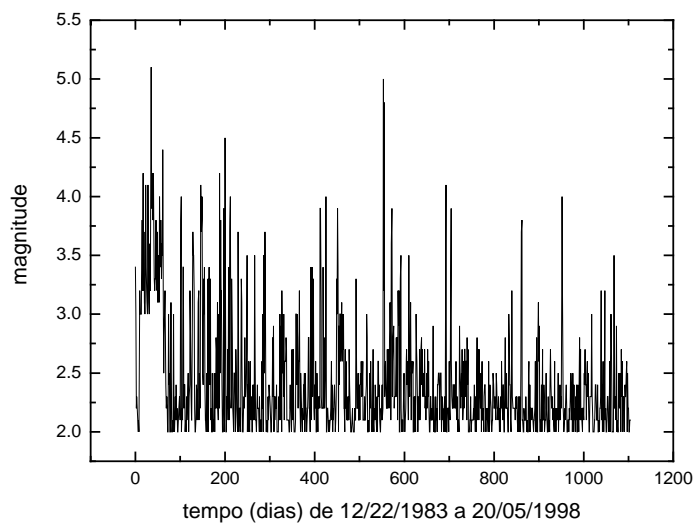

**Figura. 3: Série temporal dos dados sismológicos referentes à Falha de Samambaia de 22/12/1983 a 20/05/1998.**

Os resultados referentes à variabilidade do coeficiente de correlação apontam para a ausência de correlação. A Figura 4 mostra que a média deste coeficiente é de 0.03±0.01, portanto indicando que através deste método não existe uma correlação entre os vizinhos para qualquer defasagem $L$. Outra característica de não correlação está no rápido decaimento exponencial para pequenos valores de $L$, isto demonstra que o primeiro mínimo (ou máximo) $L_1$ (~0.85 dias) encontrado não se repete de forma sistemática para seus múltiplos ($L_1$, $2L_1$,..., $nL_1$).

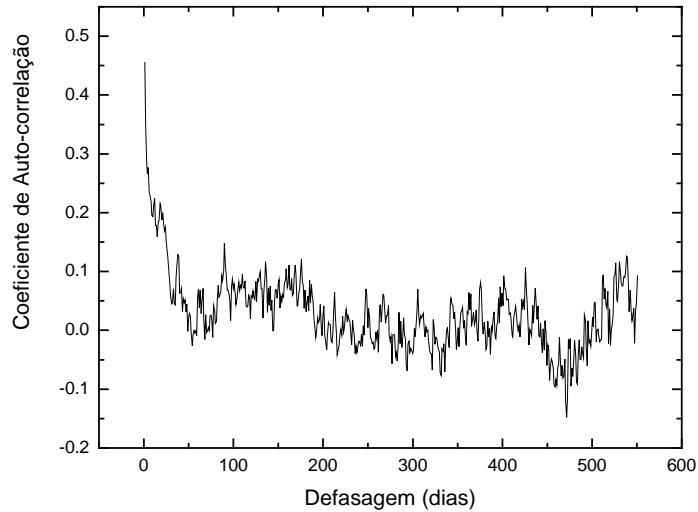

**Figura. 4: Variabilidade do coeficiente de autocorrelação dependente de escala como uma função da defasagem (em dias).**

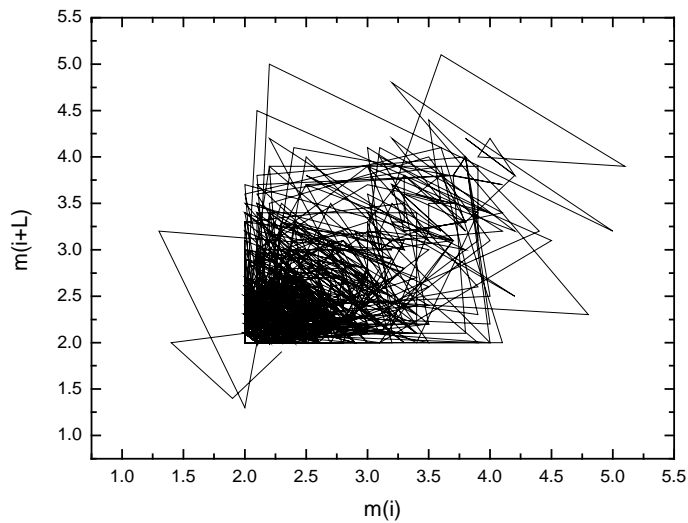

**Figura. 5: LagPlot dos dados dos terremotos para o melhor valor de L (=0.85 dias)**

Para reforçar o tratamento resolvemos verificar se este e outros valores de *L's* encontrados poderiam gerar atratores no espaço *[m(i);m(i+L)]*. A Figura 4, denominada de *LagPlot*, revela que não há formação de atrator tanto para *L*=0.85 dias quanto para qualquer outro valor de *L*. Caso houvesse surgiram curvas helicoidais indicando que este valor de *L* é uma fonte e a relação entre os vizinhos convergem para este valor.

**Expoente de Hurst**

O diagnostico da memória de longo prazo de uma série pode ser feita por meio de diversas metodologias através do cálculo do expoente de Hurst. Entre as metodologias aqui utilizadas na identificação e quantificação de memória de longo prazo, são mais utilizadas a análise R/S clássica, o método do periodograma, o método de Abry-Veitch, dentre outros especificados na Tabela 1. Como foi apresentado na Metodologia, a memória do sistema depende do valor de H para determinado intervalos previamente estabelecidos.

**Tabela 1: Resultado do expoente de Hurst dos dados sismológicos da Falha de Samambaia obtido por diferentes métodos.**

| MÉTODOS | Valor estimado de H | Coeficiente de Correlação |
|---|---|---|
| Valor absoluto | **0.832** | **0.96** |
| Variância | **0.814** | **0.99** |
| R/S | **0.348** | **0.84** |
| Periodograma | **0.923** | -- |
| Whittle | **0.77** | -- |
| Variância dos resíduos | **0.917** | **0.99** |
| Abry-Veitch | **0.766** | -- |

A Tabela 1 mostra que para todos os métodos, exceto o método R/S, o expoente de Hurst indica uma série persistente ou de memória de longa duração em contraponto ao resultado obtido pelo coeficiente de autocorrelação. Esta discrepância pode ser resultado do espaçamento da série ou a própria dinâmica de H. Este valor, obtido por cada método, representa o valor médio de H. Na verdade este valor pode variar no tempo. Além disso, vários métodos não conseguem diferenciar memória de curta duração com a de longa duração, com isso uma ou mais memória de curta duração pode ser interpretado com uma de longa duração. Para resolver este impasse poderíamos utilizar uma forma mais sofisticada de estudar a variabilidade destes expoentes utilizando análise multifractal dos dados por meio do expoente de Holder, mas isto esta fora do escopo do trabalho.

**Análise de Fourier e Wavelet**

Devido ao impasse dos métodos até então analisados resolvemos recorrer às técnicas de transformada de séries temporais. Estes métodos representam dois extremos. Na literatura a análise de Fourier é freqüentemente usada para decompor o sinal é um somatório de funções seno culminando no espectro de potência no espaço dos períodos, como representado pela Figura 6, ou no espaço das freqüências. Esta Figura apenas reforçar que o período de 0.85 dias representa um pico no espectro dos períodos mas, não representa uma assinatura de periodicidade. Com isso, deveríamos esperar que a memória de longa duração fosse representada por um pico em um determinado período seguido de seus harmônicos de várias ordens. Isto evidentemente não ocorre e, portanto, esta análise reforça a falta de memória de longa duração nesta série.

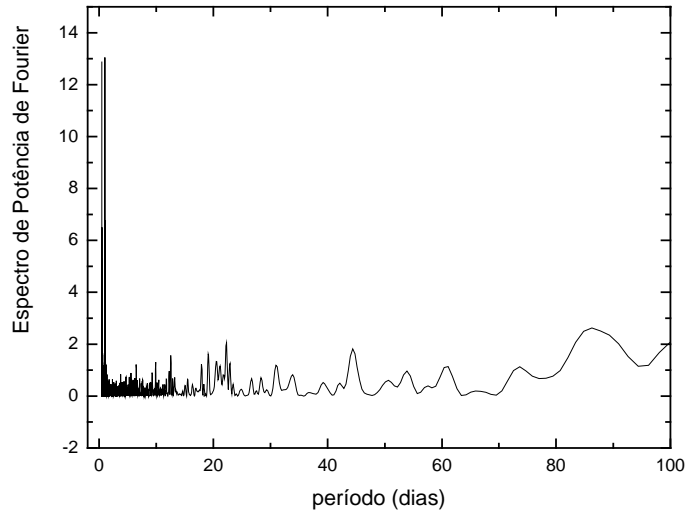

**Figura. 6: Espectro de potência de Fourier para uma janela de 100 dias.**

O outro extremo é a análise Wavelet que ao contrário da de Fourier não fixa senoídes de energia infinita mas, as colocam em pacotes gaussianos como é o caso da Wavelet aqui adotada - Wavelet de Morlet (Daubechies 1992). No caso da Wavelet o espectro de potência é representado por um mapa de curvas de nível, onde verificamos a história da variabilidade de um determinado período e as regiões mais escuras representam sua máxima variância. Como podemos verificar através da Figura 7, não existe uma periodicidade com persistência dentro da zona de confiança representada pelo cone. É evidente que um dado período ficará fora deste cone quanto mais este se aproximar do número de pontos da série. Fora deste cone existe uma periodicidade de 544.4 dias com máxima variância. Este período pode ser um harmônico de um período ainda maior que não podemos revelar, pois nossa série é apenas de poucos anos.

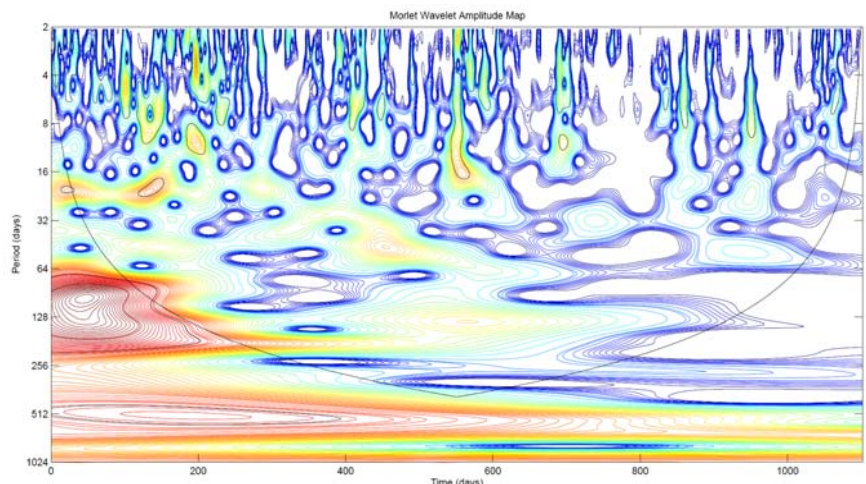

**Figura. 7: Mapa Wavelet de Morlet dos dados sismológicos da Falha de Samambaia.**

Em síntese, as análises de Fourier e Wavelet apontam para perspectivas diversas. A análise de Fourier reforça os resultados obtidos pelo coeficiente de correlação enquanto a análise Wavelet revela que existe uma persistência com baixa significância estatística podendo representa uma sinal de memória de longa duração mas, seria necessário alargar a janela temporal da amostra.

**Inferência Bayesiana**

Inicialmente, nós simulamos uma amostra que representa a opinião das pessoas baseados em uma prévia avaliação de campo sobre os possíveis fatores que as elas associam aos terremotos, principalmente, os de maior magnitude a partir do algoritmo Metropolis-Hastings. Depois desta simulação elaboramos o histograma da freqüência de ocorrência de uma dada opinião (aqui cada opinião é representada por uma variável *x* da variável aleatória *X*).

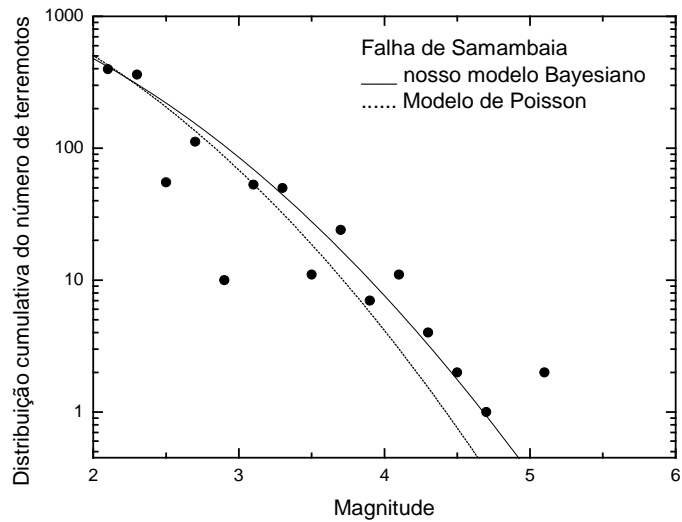

**Figura. 8: Distribuição cumulativa do número de terremotos para os dois modelos analisados.**

O próximo passo foi determinar a probabilidade de verossimilhança $l(\theta; X)$. Pela forma do histograma a função de distribuição que melhor representa é dada por:

$$l(\theta; X) = \frac{e^{-\theta X}(\theta X)^N}{N!} \qquad \text{(equação 5)}.$$

Usamos o modelo de Poisson proposto por Stavrakakis and Tselentis (1987) para comparar com o nosso de modelo Bayesiano. Para isso, recorremos ao método de regressão não-linear baseado no algoritmo de Levenberg-Marquardt para ajustar os modelos com os dados. A Figura 8 mostra a comparação entre os modelos. Como podemos verificar os modelos são equivalentes para baixas magnitudes, mas para magnitudes elevadas os modelos são crescentemente discrepantes. O modelo de Poisson possui um coeficiente de correlação de 0.45 enquanto o nosso modelo aponta para um coeficiente de 0.67. A grande diferença entre os modelos está presente na cauda da distribuição. Nela, o modelo Bayesiano encontra magnitudes compatíveis com os valores reais. Por exemplo, para o modelo de Poisson, eventos que ocorrem apenas uma vez devem ter uma magnitude em torno de 4.5 enquanto que para o nosso modelo a magnitude é de 4.7. Para ambos os modelos, magnitudes maiores que 5 não se ajustam muito, mas o modelo Bayesiano aponta para resultados mais concretos. Outra diferença entre o nosso modelo Bayesiano e o de Poisson pode ser revelado pela possibilidade de calcular a probabilidade de nenhuma ocorrência e a probabilidade de pelo menos uma em determinado tempo t (em anos). De acordo com a distribuição de probabilidade Bayesiana, nós esperamos um forte terremoto em João Câmara entre 1985 e 1987 com intervalo de magnitude entre 4.1 e 4.6, 4.7 e 5.5 ou 5.6 e 6 com probabilidade de 0.21, 0.67 e 0.15, respectivamente.

**CONCLUSSÃO**

O trabalho pode ser sumarizado em dois momentos: 1) verificar a existência de memória de longa duração nos dados sísmicos da Falha de Samambaia e 2) gerar um modelo Bayesiana que descreva com maior certeza estatística os abalos de maior magnitude ocorridos nesta falha.

Podemos concluir que os resultados apontam para interpretações diferentes sobre a questão da memória do sistema. Enquanto a função que representa a variabilidade dos coeficientes de autocorrelação, juntamente com a análise de Fourier, apontam para uma ausência de correlação o expoente de Hurst, juntamente com a análise Wavelet, apontam para uma possível correlação de longa duração nos dados sismológicos da Falha de Samambaia. Neste sentido, é necessário um estudo mais aprofundado da dinâmica que controla este fenômeno, dentre eles, já estamos iniciando um estudo sobre o formalismo multifractal baseado na transformada Wavelet para entender melhor como se estruturam as singularidades na série temporal.

Nós enfatizamos que nossos resultados, obtidos pela análise clássica (autocorrelação, Fourier, etc.) e pela Inferência Estatística, podem ser considerados como uma contribuição para o complicado problema da previsibilidade de terremotos.

**REFERÊNCIAS BIBLIOGRÁFICAS**